# Evaluation of Varrying Mobility Models & Network Loads on DSDV Protocol of MANETs

C.P.Agrawal
Computer Science Department
MCNUJC,
Bhopal, India
*agrawalcp@yahoo.com*

O.P.Vyas
*SoS in Computers
Pt Ravishankar Shukla University
Raipur, India.*
*dropvyas@gmail.com*

M.K Tiwari
*SoS in Computers
Pt Ravishankar Shukla Universit
Raipur, India*
*mtiwari91@rediffmail.com*

*Abstract* - **A mobile ad-hoc network (MANET) is collection of intercommunicating mobile hosts forming a spontaneous network without using established network infrastructure. Unlike the cellular or infrastructure networks who have a wired backbone connecting the base-station, the MANETs have neither fixed routers nor fixed locations. Their performance largely depend upon the routing mechanism & nature of mobility. Earlier research hints that the Destination Sequenced Distance Vector (DSDV) routing protocol is one of the most efficient and popular protocols, as far as general parameters have been concerned.[1,6] We have experimentally evaluated, the performance metrics for network load, packet delivery fraction and end-to-end delay with DSDV Protocol using NS2 Simulator. This paper presents, the performance of DSDV protocol for four different mobility models namely: Random Waypoint, Reference Point Group Mobility, Gauss Markov & Manhattan Mobility Model having varying network load & speed. The experimental results suggest that DSDV protocol with RPGM mobility model has optimized results for varying network load and speed.**

*Keywords: MANET, Routing Protocol, Mobility Models, Wireless Networking, DSDV protocol*

## I. INTRODUCTION

The Mobile Ad-hoc Network (MANET) is a collection of nodes, which have the possibility to connect on a wireless medium and form an arbitrary dynamic network with wireless links & without any supporting infrastructure [1,6]. The links between the nodes themselves work as routers & the network can dynamically change with time, new nodes can join, and other nodes can leave the network [3]. A MANET is of larger size than the radio range of the wireless antennas, because of this fact it can route the traffic through a multi-hop path to give two nodes the ability to communicate. There are neither fixed routers nor fixed locations for the routers as in cellular/ infrastructure networks. Cellular networks consist of a wired backbone, which connects the base-stations & the mobile nodes can only communicate over a one-hop wireless link to the base-station, multi-hop wireless links are not possible. By contrast, a MANET has no permanent infrastructure at all. All mobile nodes act as mobile routers [6].

The issue of routing packets between any pair of nodes is a challenging task because the nodes can move randomly within the network. Path that was considered optimal at a given point in time might not work a few moments later. There are two main types of the routing protocols. The proactive/ Table driven protocols maintain routes to all nodes, including nodes to which no packets are being sent, and Second approach is Demand Driven/ Source-Initiated Protocols, involves establishing reactive routes, which dictates that routes between nodes are determined solely when they are explicitly needed to route packets. With rapid improvements in channel bandwidth, speed and hardware cost of the wireless media the first approach of proactive protocols is gaining more popularity, due to their obvious better performance.

The Destination Sequenced Distance Vector (DSDV) Protocol is a proactive routing protocol, which maintain consistent, up-to-date routing information from each node to every other node in the network in a routing table at each node of the Ad-hoc network.[1, 11, 12]. These protocols require each node to maintain one or more tables to store routing information, and they respond to changes in network topology by propagating updates throughout the network in order to maintain a consistent network view. The protocol adds a new attribute, sequence number, to each route table entry at each node. Routing table is maintained at each node and with this table, node transmits the packets to other nodes in the network. Each node in the network maintains routing table for the transmission of the packets and also for the connectivity to different stations in the network. These stations list for all the available destinations, and the number of hops required to reach each destination in the routing table. The routing entry is tagged with a sequence number which is originated by the destination station. In order to maintain the consistency, each station transmits and updates its routing table periodically. The packets being broadcasted between stations indicate which stations are accessible and how many hops are required to reach that particular station. The packets may be transmitted containing the layer 2 / 3 address[4]. Routing information is advertised by broadcasting or multicasting the packets which are transmitted periodically as when the nodes move within the network. The DSDV protocol requires that each mobile station in the network must constantly, advertise to each of its neighbors, its own routing table frequently, as the entries in the table may change very quickly.

## II. MOBILITY MODELS

The mobility of nodes is one of the most important factors in the performance evaluation of MANETs. The protocol performance is highly influenced by them. The understanding of each observed mobility pattern can help to improve the network behavior. This paper is focused upon commonly used four mobility models.

### A. Random Waypoint Model

Random Waypoint Model [12, 13] is the most widely used and studied mobility model. In this model, a host randomly chooses a destination called waypoint and packet moves towards it in a straight line with a constant velocity, which is selected randomly from some given range. After it reaches the waypoint, it pauses for some time and then repeats the procedure.

For implementation in NS-2, at every instant, a node randomly chooses a destination and moves towards it with a velocity chosen randomly from $[0, V_{max}]$, where $V_{max}$ is the maximum allowable velocity for every mobile node. After reaching the destination, the node stops for a duration by the 'pause time' parameter. Then, it again chooses a random destination and repeats the whole process again until the simulation ends.

### B. Reference Point Group Mobility(RPMG) Model

Each group has a logical center (group leader) that determines the group's motion behavior. Initially, each member of the group is uniformly distributed in the neighborhood of the group leader. Subsequently, at each instant, every node has a speed and direction that is derived by randomly deviating from that of the group leader [13]. Each node deviates its velocity (both speed and direction) randomly from that of the leader. The group motion behavior is important in some applications like ubiquitous computing, military deployment etc. The movement can be characterized as follows:

$V_{member}(t) = V_{leader}(t) + random()*SDR*maxspeed$ (1)

$\theta_{member}(t) = \theta_{leader}(t) + random()* ADR*maxangle$ (2)

Where $0 <= SDR, ADR <= 1$. SDR is the Speed Deviation Ratio and ADR is the Angle Deviation Ratio. They are used to control the deviation of the velocity (magnitude and direction) of group members from that of the leader. Since the group leader mainly decides the mobility of group members, group mobility pattern is expected to have high spatial dependence for small values of SDR and ADR.

### C. Gauss-Markov Model

The Gauss-Markov Mobility Model is designed to adapt to different levels of randomness via one tuning parameter. [1,10]. Initially each mobile node (MN) is assigned a speed and direction. At fixed intervals of time $n$, movement occurs by updating the speed and direction of each MN. The value of speed and direction at the $n^{th}$ instance is calculated based upon the value of speed and direction at the $(n-1)^{th}$ instance and a random variable using equations:

$$s_n = \alpha s_{n-1} + (1-\alpha)s + \sqrt{(1-\alpha^2)}sx_{n-1} \quad (3)$$

$$d_n = \alpha dn\_1 + (1-\alpha)d + \sqrt{(1-\alpha^2)}dx_{n-1} \quad (4)$$

$s_n$ and $d_n$ are the new speed and direction of the MN at interval $n$. $\alpha$ is the tuning parameter used to vary the randomness, where $0 <= \alpha <= 1$. $s$ and $d$ are constants representing the mean value of speed and direction as $n \rightarrow \infty$ and $sx_{n-1}$ and $dx_{n-1}$ are random variables from a Gaussian distribution. Totally random values (or Brownian motion) are obtained by setting $\alpha = 0$ and linear motion is obtained by setting $\alpha = 1$. Intermediate levels of randomness are obtained by varying it between 0 and 1. At each time interval the next location is calculated based on the current location, speed, and direction of movement. Specifically, at time interval $n$, an MN's position is given by the equations:

$$x_n = x_{n-1} + s_{n-1} \cos d_{n-1} \quad (5)$$

$$y_n = y_{n-1} + s_{n-1} \sin d_{n-1} \quad (6)$$

Where $(x_n, y_n)$ and $(x_{n-1}, y_{n-1})$ are the $x$ and $y$ coordinates of the MN's position at $n^{th}$ and $(n-1)^{th}$ time intervals respectively. To ensure that a MN does not remain near an edge of the grid for a long period of time, the they are forced away from an edge when they move within a certain distance of the edge. This Model can eliminate the sudden stops and sharp turns encountered in the Random Walk Mobility Model by allowing past velocities to influence future velocities.

### D. Manhattan Mobility Model

The Manhattan model emulates the movement pattern of mobile nodes on streets defined by maps [12,13]. It is useful in modeling movement in an urban area where a pervasive computing service between portable devices is provided. Maps are composed of a number of horizontal and vertical streets used in this model. Each street has two lanes for each direction (North / South direction for vertical streets, East / West for horizontal streets). The mobile node is allowed to move along the grid of horizontal and vertical streets on the map. At an intersection of a horizontal and a vertical street, the mobile node can turn left, right or go straight. This choice is probabilistic, the probability of moving on the same street is 0.5, the probability of turning left is 0.25 and the probability of turning right is 0.25. The velocity of a mobile node at a time slot is dependent on its velocity at the previous time slot. Also, a node's velocity is restricted by the velocity of the node preceding it on the same lane of the street..

## III. SIMULATION SCANERIO

Simulations have been carried out by Network Simulator 2.27 NS-2 [8, 9]. We have used an average packet size of 350 bytes at a varying rate of 4, 8, 12 and 16 packets/s taken uniform 100 nodes with constant pause time 10s [5]. In this simulation we wanted to investigate how the protocol behaves when network load and speed of nodes increases with different Mobility Models.

Continuous bit rate (CBR) traffic sources are used. The source-destination pairs are spread randomly over the network. The number of source-destination pairs is constant and the packet-sending rate in each pair is varying. The four traffic and mobility models taken for the experiments are in a rectangular field of size 500 m x 500 m. For Random Waypoint each packet starts its journey from a random location to a random destination. Once the destination is reached, another random destination is targeted after a pause. The pause time, which affects the relative speeds of the mobiles, is kept constant at 10 s. Simulations are run for 100 seconds. Identical mobility and traffic scenarios are used across protocols to gather fair results. Rest movement pattern are according to the characteristics of RPGM Model and Gauss Markov Mobility Model.

For scenarios generation we have used Bonn Motion, a mobility scenario generation and analysis tool. Bonn Motion is a Java software, which creates and analyses mobility scenarios. It serves as a tool for the investigation of mobile ad hoc network characteristics. The scenarios can also be exported for the network simulator ns-2 and GlomoSim / QualNet . In the simulation scenario, the three input parameters have been  varied to evaluate the output performance.  Parameters values  are defined below.

- The speed of  nodes – It has been varied from 5 to 25 m/s (on X axis).

- Network Load- Number of packets sent per second has been varied from 4 to 16

packets/s. Four different graphs have been plotted for this parameter.

- Mobility Model: The graphs for each of the four mobility models have been plotted on every plot.

- The simulation results for two output parameters have been observed from the NS-2 results for plotting graphs using Excel for the observed output parameter values.

- Average Delay- This average End-to-End delay includes processing and queuing delay in each intermediate node. This parameter has been indicated on Y axis, for the set (A) of the graphs.

- Packet Delivery fraction (PDF)- Packet delivery fraction is calculated as the ratio of the number of packets received (by the destination) to the number of packets originated by the (CBR source). This has been indicated on Y axis, for the set (B) of graphs.

TABLE I. PARAMETERS CHOSEN FOR SIMULATION IN NS-2

| Parameter | Value |
|---|---|
| Simulation Time | 100 sec. |
| Number of Nodes | 100 |
| Pause Time | 10 sec. |
| Environment Size | 500 m X 500m |
| Traffic Type | Constant Bit Rate |
| Maximum Speeds | 5, 10, 15, 20, 25 m/s |
| Network Loads | 4, 8, 12, 16 packets/s |
| Mobility Models | Random Waypoint Model, Reference Point Group Mobility (RPGM) Model, Gauss Markov Model and Manhattan Models. |

IV. GRAPHICAL REPRESENTATION OF SIMULATION RESULTS

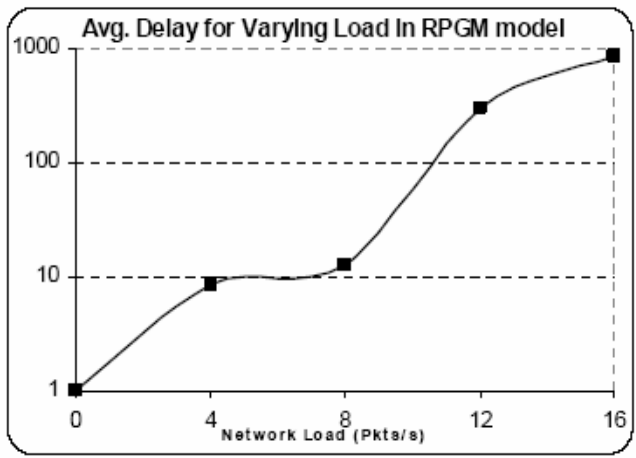

FIGURE – 1 (A)

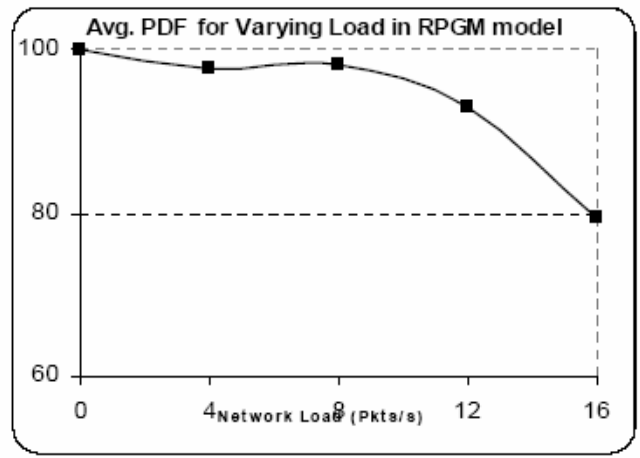

FIGURE – 1 (B)

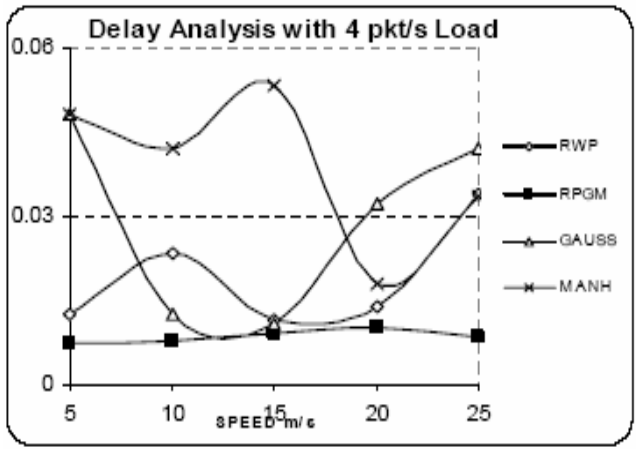
FIGURE – 2 (A)

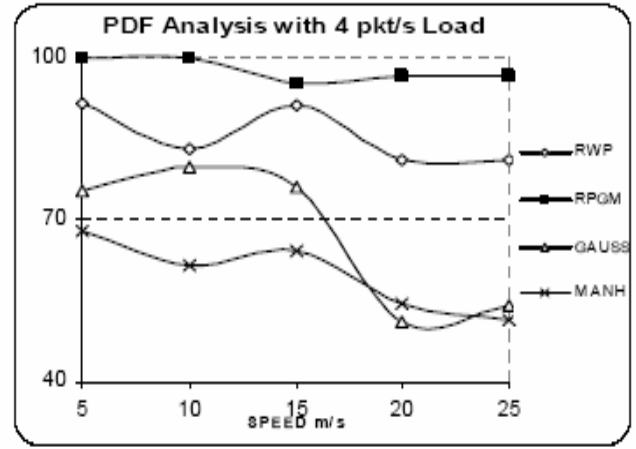
FIGURE – 2 (B)

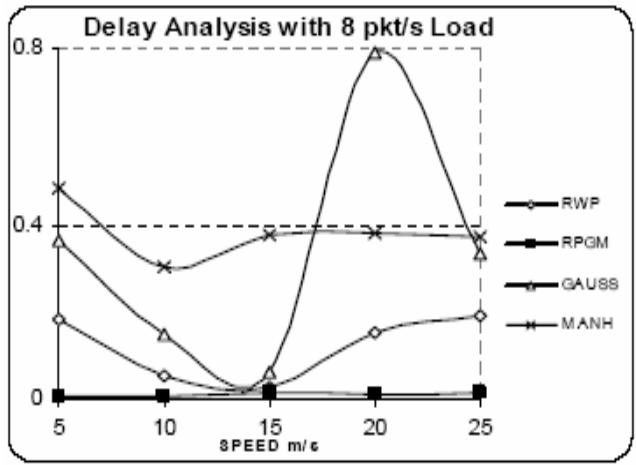
FIGURE – 3 (A)

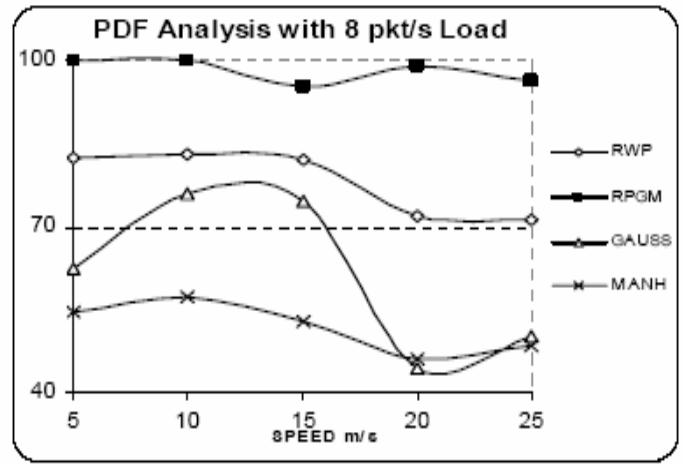
FIGURE – 3 (B)

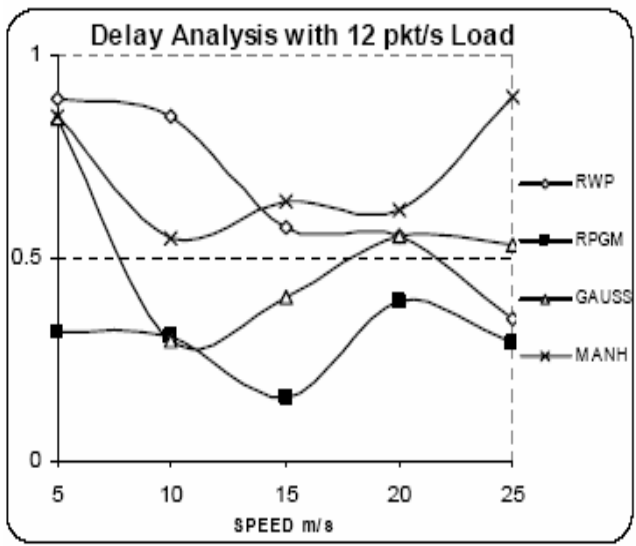
FIGURE – 4 (A)

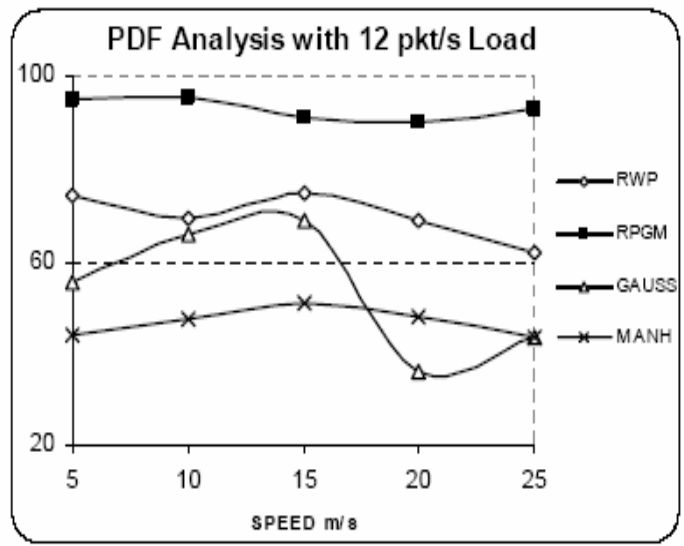
FIGURE – 4 (B)

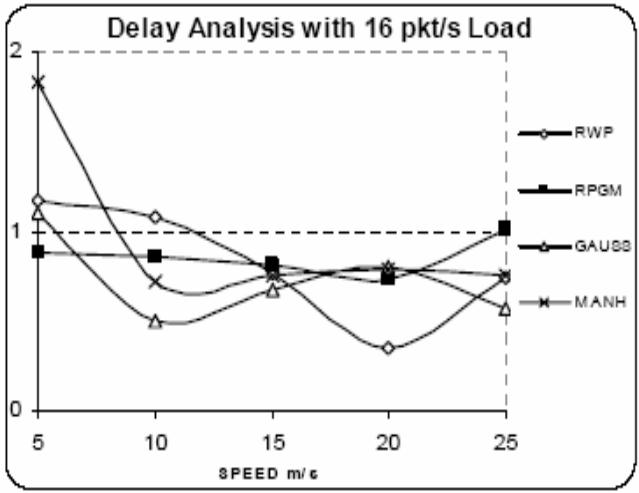
FIGURE – 5 (A)

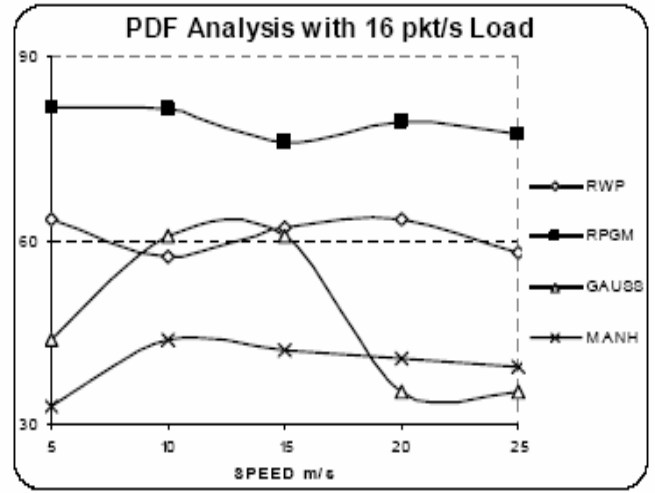
FIGURE – 5 (B)

## V. RESULTS & DISCUSSION

The results of investigation obtained as Packet Delivery Fraction (PDF) and Average end-to-end Delay for DSDV protocol, with the four mobility models for 5, 10, 15, 20, 25 m/s speed of nodes have been indicated and compared for different 4, 8, 12 & 16 packets/s network loads for 100 nodes. In each of the graphs, Figures 1 through 4, the RPGM Mobility model gave the best performance (shown with filled squares). The curve for average of all delays against the varying load has only been plotted for this best performing mobility model.

The overall simulation experiments results suggest that in the considered simulation scenario at increasing network load and speed of nodes, selecting DSDV with RPGM Mobility Model would be best in order to have higher delivery of packets with lowest delay. The average of all end-to-end delays for different speeds has been found increasing exponentially (drawn on logarithmic scale on Y axis) with respect to the Network Load [Figure 1(A)]. On the other hand the average of all PDFs for different speeds has been found decresing with respect to the Network Load [Figure 1(B)]. The DSDV with Manhattan exhibited worst performance at each of the network scenario as compared to the other three mobility models.

The performance of MANETs is largely dependent on the pattern of mobility of the moving nodes. Our efforts of performance optimization of DSDV protocol yields in improvement under varying network load and also for different mobility models.

The outputs experimentally obtained, suggest that DSDV with RPGM mobility model exhibits best performance in terms of PDF and end-to-end delay. Thus the performance of RPGM Mobility Model is better selection keeping other scenarios constant. This is theoretically justified, due to the fact that each group has a logical center or group leader that determines the group motion behavior. Each member of group is uniformly distributed in the neighborhood of the group leader and each node deviates its velocity both speed and direction randomly from that of leader.

The experimentation also suggests that several parameters such as traffic patterns, node density and initial pattern of nodes also affect the routing performance and need to be investigated with various scenarios. Further study also needs to be done with additional analysis of mobility models with different Adhoc Routing Protocols for varying mobility, network load and pause-time for better evaluation of performance of these mobility models and Adhoc Networks protocols.

### A. For 4 Pkt/s Network Load

The RPGM mobility model has performed best as compared to other mobility models. The delay is between 7 & 10 ms [Figure 2(A)] and the PDF between 95 & 100% [Figure 2(B)]. Mostly the delay is found increasing with the speed of nodes. Performance for Manhattan mobility model has been the poorest performance.

### B. For 8 Pkt/s Network Load

The RPGM mobility model has performed best as compared to other mobility models. The delay is between 9 & 16 ms [Figure 3(A)] and the PDF between 95 & 100% [Figure 3(B)]. Mostly the delay is found increasing with the speed of nodes. Performance for Manhattan mobility model has been the poorest performance.

### C. For 12 Pkt/s Network Load

With the increased network load to 12 Pkt/s also the RPGM mobility model has performed best as compared to other mobility models. The delay is between 158 & 393 ms [Figure 4(A)] and the PDF between 89 & 95% [Figure 4(B)]. Performance for the other three mobility models have been fluctuation as for the two output parameter comparison. Broadly the Manhattan mobility model has been the poorest performance.

### D. For 16 Pkt/s Network Load

With the highest network load of our simulation experiment for 16 Pkt/s also the RPGM mobility model has performed best as compared to other mobility models. The delay is between 0.727 sec & 1 sec [Figure 4(A)] s and the PDF between 76 & 81%[Figure 4(B)]. Actually for the delay, readings for the other three mobility models have shown fluctuating performance at different speeds, even going below the RPGM model ratings at some speed values. But the overall delay for RPGM has been consistently the best if all speed values are considered. Again the Manhattan mobility model has been the poorest performance.

## VI. FUTURE WORK

This work was mainly relating to the popular mobility models. Similar results for other popular MANET routing protocols DSR and AODV , have also been published by the same authors in international conferences [2, 4]. As a natural outcome and extension to this work, further analysis needs to be carried out to compare the three sets, so as to cumulatively infer more complete conclusions.